\documentclass[%
 reprint,
 superscriptaddress,
 amsmath,amssymb,
 aps,
 pra,
floatfix,
]{revtex4-2}
\usepackage{graphicx} 
\usepackage{dcolumn} 
\usepackage{bm} 
\usepackage[english]{babel}
\usepackage{physics}
\usepackage[utf8]{inputenc}
\usepackage{amsthm}
\usepackage{mathtools}
\usepackage{physics}
\usepackage{xcolor}
\usepackage{graphicx}
\usepackage[left=15mm,right=15mm,top=25mm,columnsep=15pt]{geometry} 
\usepackage{adjustbox}
\usepackage{placeins}
\usepackage[T1]{fontenc}
\usepackage{lipsum}
\usepackage{csquotes}
\usepackage{float}
\usepackage{hyperref}
\usepackage{enumitem}
\usepackage{appendix}
\usepackage{soul}
\usepackage{physics}
\usepackage{braket}
\usepackage{esvect}
\usepackage{amsmath}

\usepackage{algpseudocode}
\usepackage{algorithm}
\usepackage{multirow}
\usepackage{lineno}

\usepackage[color=yellow]{todonotes}
\begin{document}

\title{Trapped ion quantum hardware demonstration of energy calculations using a multireference unitary coupled cluster ansatz: application to the BeH$_2$ insertion problem}

\author{Palak Chawla}
\altaffiliation{Contributed equally to the work}
\affiliation{Centre for Quantum Engineering, Research and Education, TCG Crest, Kolkata 700091, India}
\author{Disha Shetty}
\altaffiliation{Contributed equally to the work}
\affiliation{Centre for Quantum Engineering, Research and Education, TCG Crest, Kolkata 700091, India}
\author{Peniel Bertrand Tsemo}
\affiliation{Centre for Quantum Engineering, Research and Education, TCG Crest, Kolkata 700091, India}
\affiliation{Department of Physics, IIT Tirupati, Chindepalle, Andhra Pradesh 517619, India}
\author{Kenji Sugisaki}
\affiliation{Centre for Quantum Engineering, Research and Education, TCG Crest, Kolkata 700091, India}
\affiliation{Graduate School of Science and Technology, Keio University, 7-1 Shinkawasaki, Saiwai-ku, Kawasaki, Kanagawa 212-0032, Japan}
\affiliation{Quantum Computing Center, Keio University, 3-14-1 Hiyoshi, Kohoku-ku, Yokohama, Kanagawa 223-8522, Japan}
\affiliation{Keio University Sustainable Quantum Artificial Intelligence Center (KSQAIC), Keio University, 2-15-45 Mita, Minato-ku, Tokyo, Japan}
\author{Jordi Riu}
\affiliation{Qilimanjaro Quantum Tech, Carrer de Veneçuela, 74, Sant Martí, 08019, Barcelona, Spain}
\affiliation{Universitat Politècnica de Catalunya, Carrer de Jordi Girona, 3, 08034 Barcelona, Spain}
\author{Jan Nogué}
\affiliation{Qilimanjaro Quantum Tech, Carrer de Veneçuela, 74, Sant Martí, 08019, Barcelona, Spain}
\affiliation{Universitat Politècnica de Catalunya, Carrer de Jordi Girona, 3, 08034 Barcelona, Spain}
\author{Debashis Mukherjee}
\affiliation{Centre for Quantum Engineering, Research and Education, TCG Crest, Kolkata 700091, India}
\author{V. S. Prasannaa}
\email{srinivasaprasannaa@gmail.com}
\affiliation{Centre for Quantum Engineering, Research and Education, TCG Crest, Kolkata 700091, India}
\affiliation{Academy of Scientific and Innovative Research (AcSIR), Ghaziabad- 201002, India} 

\begin{abstract}
In this study, we employ the variational quantum eigensolver algorithm with a multireference unitary coupled cluster ansatz to report the ground state energy of the BeH$_2$ molecule in a geometry where strong correlation effects are significant. We consider the two most important determinants in the construction of the reference state for our ansatz. We remove redundancies in order to execute a redundancy-free  calculation. In view of the currently available noisy quantum hardware, we carry out parameter optimization on a classical computer and measure the energy with optimized parameters on a quantum computer. Furthermore, in order to carry out our intended 12-qubit computation with error mitigation and post-selection on a noisy intermediate scale quantum era trapped ion hardware (the commercially available IonQ Forte-I), we perform a series of resource reduction techniques to  a. decrease the number of two-qubit gates by $99.84\%$ (from 12243 to 20 two-qubit gates) relative to the unoptimized circuit, and b. reduce the number of measurements via the idea of supercliques, while losing $2.69\%$ in the obtained ground state energy relative to that computed classically for the same resource-optimized problem setting. 
\end{abstract} 

\maketitle

\section{Introduction}

The field of quantum chemistry using quantum computers has been one of the most rapidly emerging areas of research in quantum sciences and technologies, owing to the promise of the speedup that quantum algorithms offer for solving such problems \cite{Abrams1997sim, Abrams1999QPE, kitaev1995quantum, aspuru2005simulated}. On the quantum computing front, since we currently are in the noisy intermediate scale quantum (NISQ) era \cite{preskill2018quantum}, calculations involving over a handful of qubits are primarily limited by the currently achievable two-qubit gate fidelities. In such a scenario, despite the limited number of qubits available, it is important to push the boundaries of quantum chemical computations using quantum computers, while maximizing both precision and the number of qubits that can be effectively utilized. On the other hand, on the quantum chemistry front, the quest to accurately predict molecular energies in strongly correlated regimes has long been an issue of fundamental importance owing to the variety of applications that it can find \cite{laidig1987description, bulik2015can, evangelista2018perspective, manna2019simplified, meissner2004breaking, kaldor1989multireference}.  

In this work, we aim to carry out a quantum chemical calculation where strong correlation effects are at play on a NISQ era quantum computer. To that end, we use the widely employed quantum-classical hybrid Variational Quantum Eigensolver (VQE)\cite{Peruzzo2014VQE, o2016scalable, kandala2017hardware,  kawashima2021optimizing, google2020hartree, nam2020ground, mccaskey2019quantum, hempel2018quantum, colless2018computation, yamamoto2022quantum, Jules2022VQE_review, Pan2023vqe12q, google2023purification} algorithm with a multireference unitary coupled cluster ansatz (MRUCC-VQE). We remark at this point that several works exist in literature to capture strong correlation effects using the VQE algorithm \cite{halder2022dual, wu2024multi, greene2021generalized, lee2018generalized, guo2024experimental, sugisaki2022variational}. However, while many studies focus on the extremely important formulation step, few take the next step of implementing and executing a problem instance on a quantum computer, where the key challenges lie in reducing circuit depth while retaining precision to as much extent as possible. To the best of our knowledge, only one work employs a MRUCC-VQE approach \cite{guo2024experimental} on quantum hardware. The work uses a superconducting qubit quantum computer to obtain energies of some light molecular systems. Our work is based on the multireference ansatz discussed in Sugisaki \textit{et al} \cite{sugisaki2022variational}, where the authors simulate the BeH$\mathrm{_2}$ insertion problem in the MRUCC-VQE framework. We carry out a 12-qubit computation (with error mitigation) on a commercially available trapped ion quantum computer for the same problem. Among the commercially available quantum computers, trapped ion devices offer among the best two-qubit gate fidelities as well as all-to-all connectivity, thus making the platform our choice for this work. In view of steep costs involved in executing a task on a good quality quantum computer and due to error accumulation in our final result that occurs in an iterative process, we only carry out the energy estimation step on quantum hardware with the parameter optimization process is simulated on a traditional computer. 

The manuscript is organized as follows: We discuss the BeH$\mathrm{_2}$ insertion problem (Section \ref{beh2}) followed by details of MRUCC-VQE (Section \ref{icmrucc}) and then proceed to expound on our resource reduction techniques (Section \ref{resred}). Section \ref{results} discusses our results obtained on the commercially available IonQ Forte-I quantum computer. We note that all our energies are rounded off to six decimal places.

\section{Background and our workflow}\label{Sec2} 

\subsection{The BeH$\mathrm{_2}$ insertion problem} \label{beh2} 
\begin{figure*}[t]
\begin{tabular}{cc}
\includegraphics[scale=0.2]{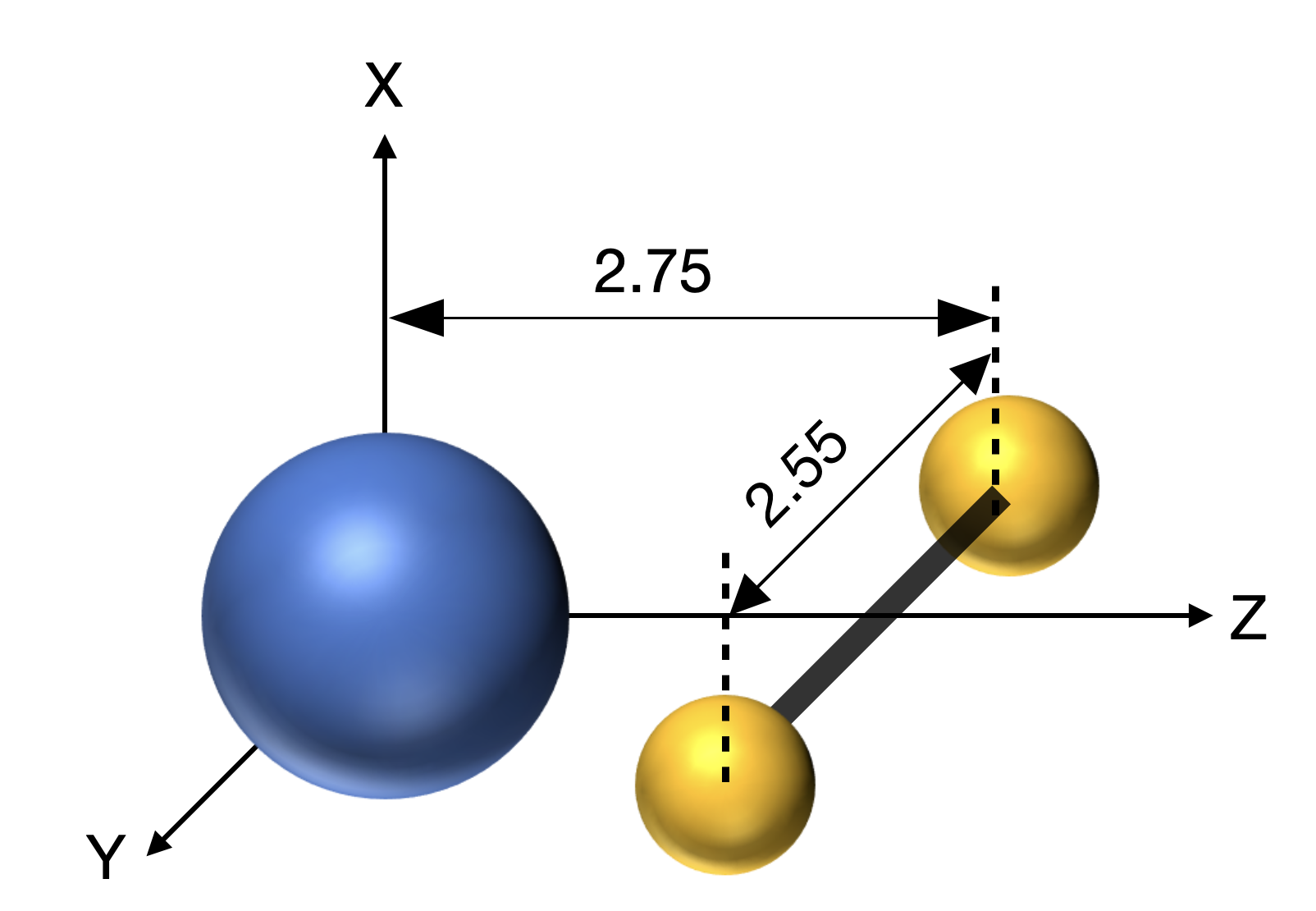}&\includegraphics[scale=0.2]{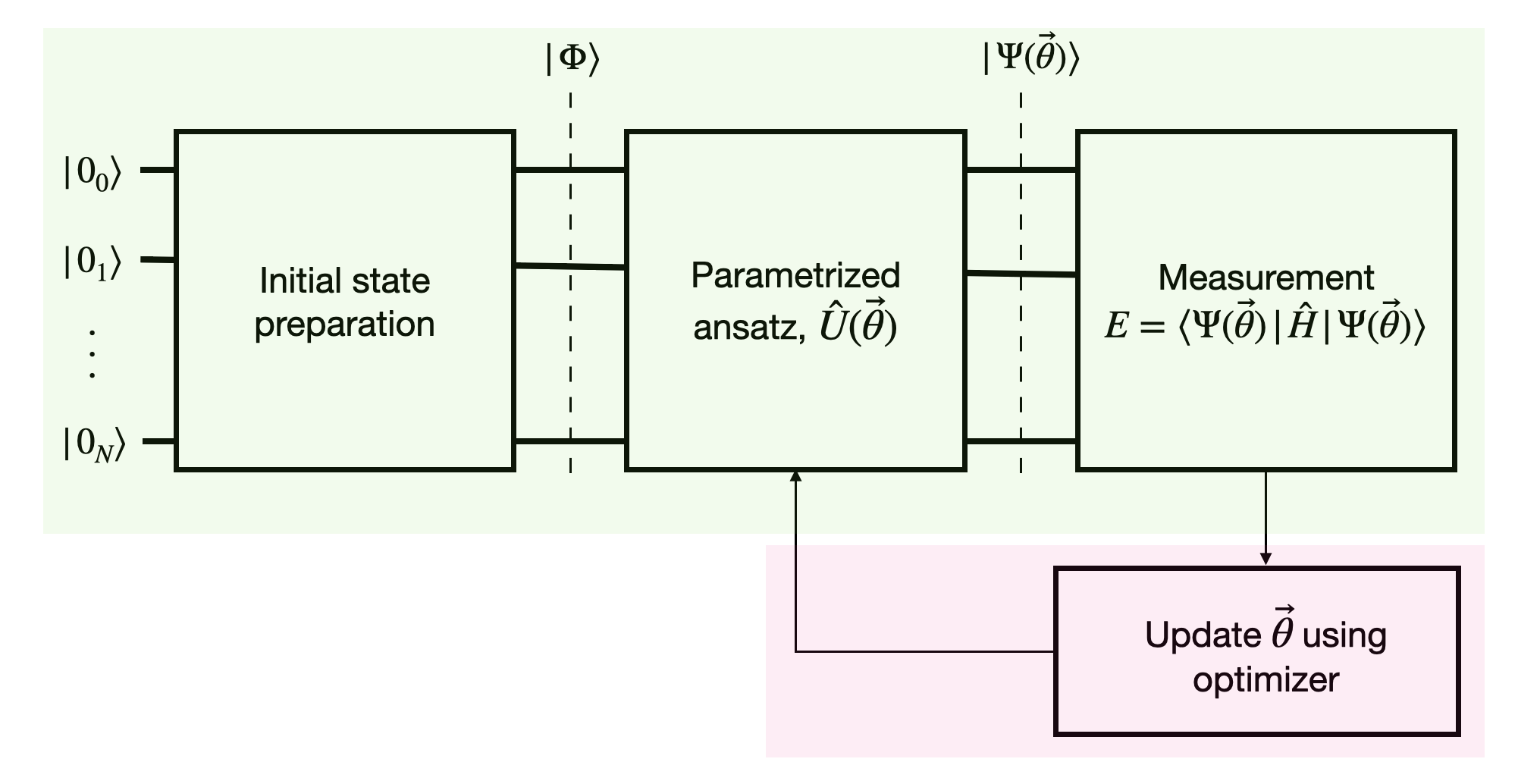}\\
(a) & (b)\\
\includegraphics[scale=0.5]{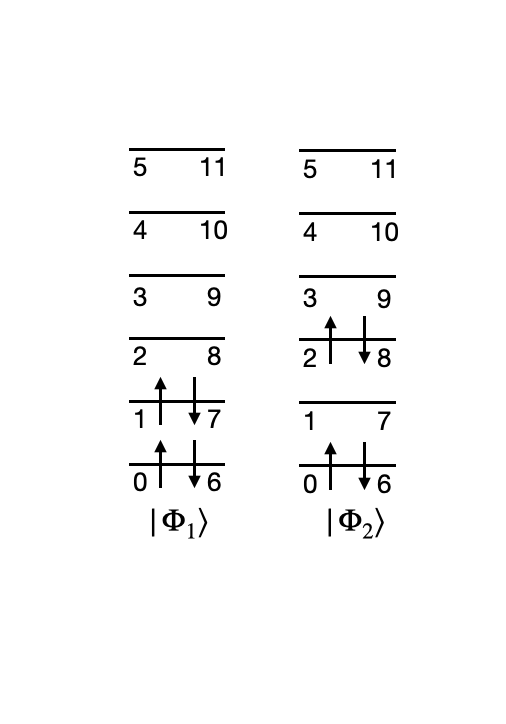}&
\includegraphics[scale=0.3]{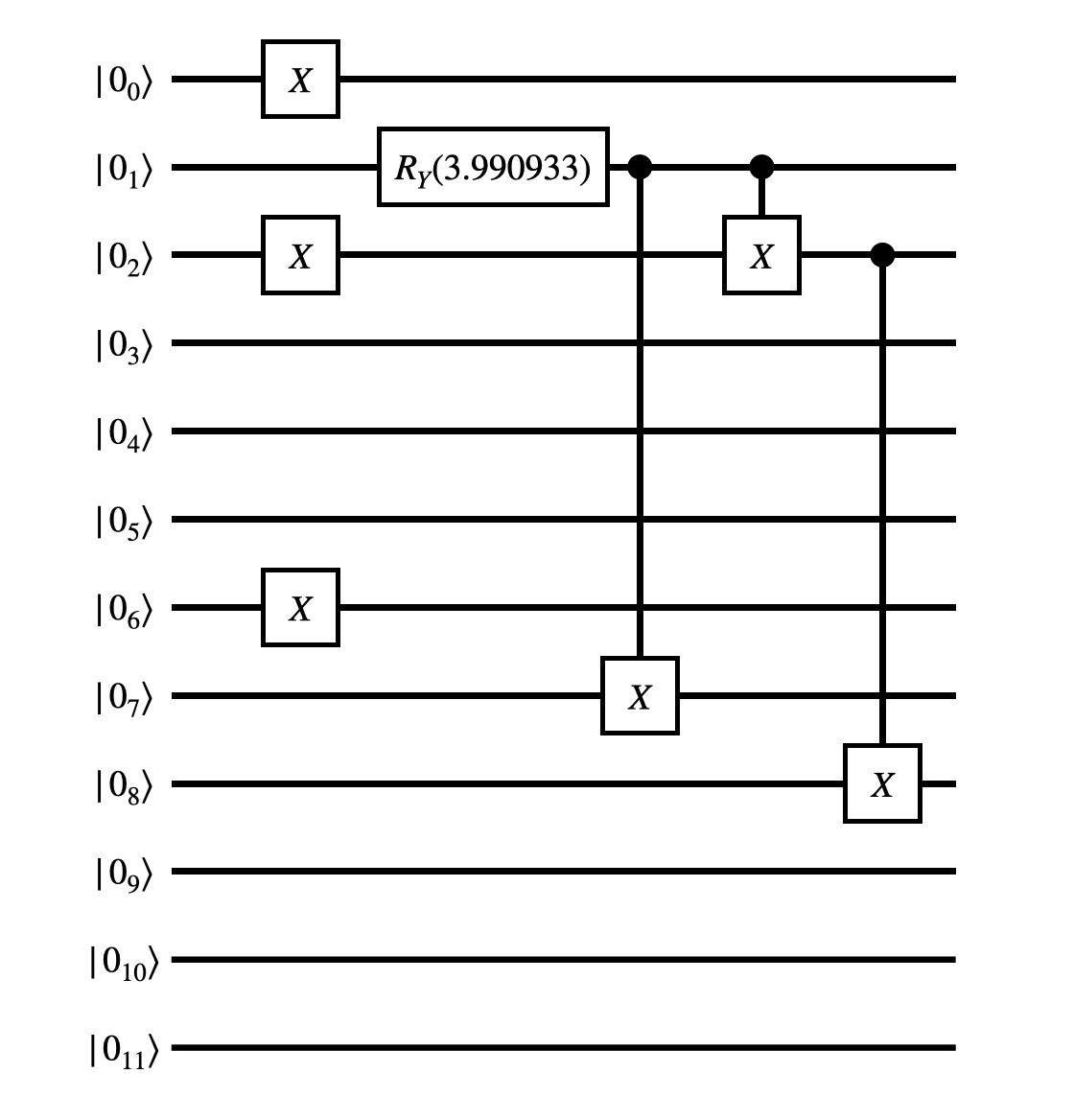}\\ 
(c)&(d)\\ 
\end{tabular}
\caption{(a) A schematic of the BeH$_2$ insertion geometry that we consider for this work (not to scale) with units in Bohr, (b) an outline of the VQE algorithm, (c) an orbital diagram representation of the two reference determinants considered, and (d) the quantum circuit to prepare the initial state for an MRUCCSD-VQE calculation, $0.911174\ket{110000110000}-0.412020\ket{1010001010000}$, starting from the fiducial state $\ket{0}^{\otimes12}$. } 
\label{fig:1}
\end{figure*}

We consider the BeH$\mathrm{_2}$ insertion problem~\cite{BeH21}. The problem involves a pathway, where the Be atom moves towards the H$_2$ molecule in a direction perpendicular to the H$_2$ molecule's bond axis, and breaks the existing bonding orbital between the two hydrogen atoms, leading to the formation of a bonding orbital between the $2p_z$ orbital of Be and the anti-bonding orbital, $\sigma_u$, of H$_2$. The bond length at which this `insertion' happens is the one that is of interest to us in this work. The coordinates of the two Hydrogen atoms that we consider are $x=0.000, y=\pm 1.275$ and $z=2.750$, with the units given in Bohr (see Figure \ref{fig:1}(a) for a schematic, drawn not to scale). The Be atom is placed at the origin. We work with the STO-6G basis, and use the $C_{2v}$ point group symmetry in our computations. Our choice for the geometry corresponds to the point along the well-studied Be $+$ H$\mathrm{_2}$ $\rightarrow$ BeH$_2$ reaction pathway, where an avoided crossing occurs between the first and the second singlet states of BeH$_2$ in $A1$ symmetry. At this geometry,  multireference effects are pronounced, and thus the single reference unitary coupled cluster ansatz would not yield good quality results, thus necessitating the use of a multireference coupled cluster treatment~\cite{sugisaki2022variational}. 

\subsection{The MRUCC framework} \label{mrucc}

We now expand on the last part of the previous sub-section. Multi-reference effects are said to be important when the Hartree-Fock (HF) reference state is no longer a good starting point. This is due to the fact that more than one determinant in the configuration interaction (CI) wave function expansion becomes important in the sense of their accompanying coefficients contributing significantly. For example, in the problem that we consider, two determinants carry heavy weights in its full CI expansion: 0.724 and 0.560 for $1a_1^22a_1^21b_2^2$ (pertaining to HF) and the $1a_1^22a_1^23a_1^2$ configurations respectively \cite{BeH21}. This can be contrasted with, for example, the H$_2$ molecule in the STO-3G basis (and in the equilibrium geometry), where the HF determinant carries a weight of $0.9939$ while the second most important configuration (a doubly excited determinant) has a weight of $-0.1106$ \cite{helgaker2013molecular}. 

The coupled cluster (CC) wave function for an $N_e$-electron system is described as $\ket{\Psi} = e^{\hat{T}}\ket{\Phi_0}$, where $\hat{T} = \hat{T}_1 + \hat{T}_2 + \cdots + \hat{T}_{N_e}$, with $\hat{T}_1=\sum_{ia}t_{ia}\hat{a}_a^\dag \hat{a}_i$, $\hat{T}_2 = \sum_{ijab} t_{ijab} \hat{a}_a^\dag \hat{a}_b^\dag \hat{a}_i \hat{a}_j$, etc. We use the convention $i,j,\cdots$ for occupied and $a,b,\cdots$ for virtual spinorbitals. The unitary version of the CC method, aptly named the unitary coupled cluster (UCC) method, describes the wave function as $\ket{\Psi} = e^{\hat{\tau}}\ket{\Phi_0};\ \hat{\tau} = \hat{T} - \hat{T}^\dag$, and is best suited for quantum computing implementations (for example, see Ref. \cite{mcardle2020quantum}). Several multi-reference CC (MRCC) formalisms exist \cite{Mahapatra1998, Mukherjee01121975, Monkhorst1981, Paldus1988}, but for the purposes of this work, we pick the straightforward ansatz $\hat{U}(\vec{\theta})\ket{\Phi} = e^{\hat{\tilde{\tau}}}\ket{\Phi}=e^{\hat{\tilde{\tau}}}\sum_{i}C_i\ket{\Phi_i}$. In the above expression, the reference state has been expanded as a linear combination of determinants. $\hat{\tilde{\tau}}$ is built out of linearly independent excitations. It is necessary to use this operator on the exponent in place of the standard UCC operator (the t-amplitudes are the parameters out of which $\vec{\theta}$ is built), in order to ensure that there are no redundancies. Redundancies occur when an excitation operator's action on a determinant from the set $\{\ket{\Phi_i}\}$ and another excitation operator's action on another such determinant lead to the same output determinant. For example, in the case of BeH$_2$ that we consider, we see that the action of the operator that leads to the double excitation $(0 \rightarrow 2, 1 \rightarrow 4)$ on $\ket{110000110000}$ and the action of that which leads to the double excitation $(0 \rightarrow 4, 8 \rightarrow 7)$ on $\ket{101000101000}$ yield to the same final excited determinant, $\ket{001010110000}$. In this work, we manually remove the redundancies, as we shall explain in a subsequent section. 

\subsection{MRUCC-VQE algorithm} \label{icmrucc} 


We present a brief description of the VQE algorithm. It is a quantum-classical hybrid approach that relies on the Rayleigh-Ritz variational principle ~\cite{gould1966vqe}. The algorithm involves minimizing an energy functional $E(\vec{\theta}) = \frac{\expval{\hat{H}}{\Psi(\vec{\theta})}}{\braket{\Psi(\vec{\theta})|\Psi(\vec{\theta})}} = \langle \Phi |\hat{U}^\dag(\vec{\theta}) \hat{H} \hat{U}(\vec{\theta})|\Phi \rangle$ with respect to the parameters $\{\vec{\theta}\}$ via an iterative procedure, where by starting with an initial guess for the parameters, one updates them at each iteration using an optimizer routine. The parameter update step is carried out on a classical computer whereas the energy functional evaluation for each iteration is done on a quantum computer (see Figure \ref{fig:1}(b) for a schematic describing the algorithm)~\cite{Peruzzo2014VQE}. Here, $\hat{H}$ is the molecular Hamiltonian given by $\sum_{pq} h_{pq} \hat{a}_p^\dagger \hat{a}_q + \frac{1}{2} \sum_{pqrs} h_{pqrs} \hat{a}_p^\dagger\hat{a}_q^\dagger \hat{a_s}\hat{a_r}$ in the second quantized form, where $h_{pq}$ and $h_{pqrs}$ are the one- and two-electron integrals, while $p, q,\cdots$ denote spinorbital indices (occupied and unoccupied). Each index is summed over the total number of spinorbitals, $N$. $\ket{\Psi(\vec{\theta})}$ is the molecular wave function, and is expressed as a unitary $\hat{U}(\vec{\theta})$ acting on a reference state $\ket{\Phi}$, where $\hat{U}(\vec{\theta})$ is described by the MRUCC ansatz specified in the earlier sub-section. In this work, we only consider single (S) and double (D) excitations, thus the ansatz in use is MRUCCSD.

\begin{figure*}[t]
\setlength{\tabcolsep}{0.6mm}
        \begin{tabular}{cc}
        \hspace{-0.6cm}
            \centerline{\includegraphics[scale=0.6]{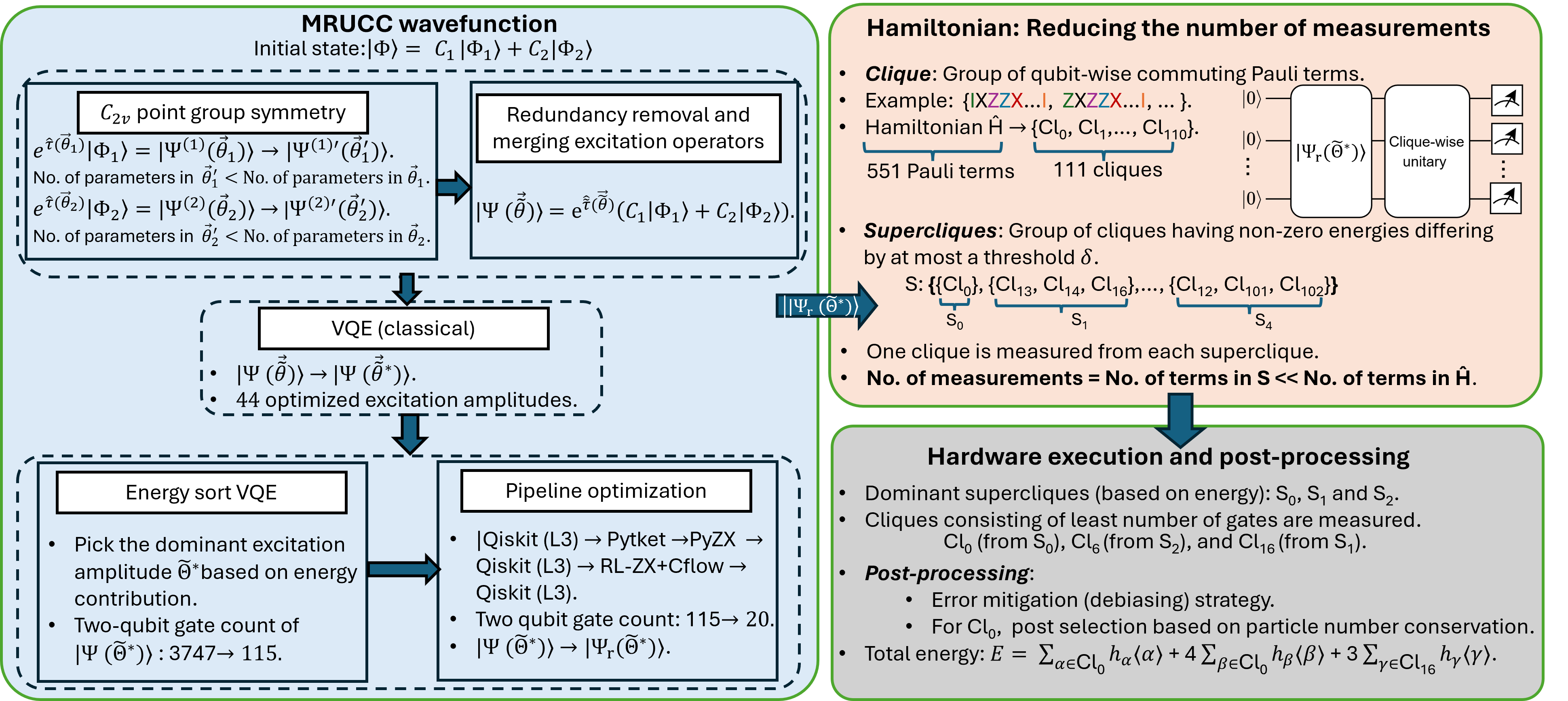}}
            \\
            \hspace{0.3cm}\\
    
        \end{tabular}
\caption{An illustration of the workflow adapted in our work, consisting of optimizations carried out at the wavefunction level (left panel) and at the Hamiltonian level (right panel). Left panel: $e^{\hat{\tau}{(\vec{\theta}_1)}}$ is the UCC operator acting on the first determinant, $\ket{\Phi_1}$, whereas $e^{\hat{\tau}{(\vec{\theta}_2)}}$ acts on the second determinant, $\ket{\Phi_2}$. Upon using point group symmetry, the number of non-zero amplitudes are reduced. Thus, $\vec{\theta}_1'$, for example, is $\vec{\theta}_1$ with several elements in the latter zeroed out. $\hat{\tilde{\tau}}$ refers to the UCC operator that contains only the linearly independent amplitudes upon removing redundancies and merging identical excitation operators. In the energy sort VQE sub-panel, the dominant optimized parameter that we pick is expressed as $\tilde{\Theta}^*$. The reduced state, $\ket{\Psi(\tilde{\Theta}^*)}$, is passed through a pipeline optimization routine to get $\ket{\Psi_r(\tilde{\Theta}^*)}$ . Right panel: the Hamiltonian, $\hat{H}$, is partitioned into qubit-wise mutually commuting sets (shown with colour coding) called cliques (denoted as Cl$_0$ for the $0^{th}$ clique, etc). The cliques are grouped under supercliques based on a set threshold $\delta$. We pick the top three supercliques for our calculation. The expectation value is then calculated (with error mitigation) using quantum hardware for the reduced Hamiltonian, with respect to this pipeline optimized wavefunction to obtain the counts. We then pass the counts obtained from clique 0 through a post-selection step to obtain the final set of counts, and thus the final energy. We note that IXZZX$\cdots$I is a shorthand for $\mathrm{\hat{I}\otimes \hat{X}\otimes \hat{Z}\otimes \hat{Z}\otimes \hat{X}\otimes \cdots \otimes \hat{I}}$. } 
\label{fig:2}
\end{figure*} 

\subsection{Our workflow} 

We plan to carry out a 12-qubit VQE computation (which corresponds to a 12-spinorbital calculation). The 14-qubit computation reduces to a 12-qubit one upon freezing the innermost molecular orbital. We now briefly comment on how the quantum circuit is executed. The expression presented in the previous paragraph for the state, $\ket{\Psi(\vec{\theta})}$, is recast into its quantum circuit form by using Jordan-Wigner transformation \cite{Jordan1928, seeley2012bravyi}, to convert fermionic operators into qubit operators, followed by Trotterization and finally applying Pauli gadgets (for example, see Refs.\cite{whitfield2011simulation, mcardle2020quantum}). The coefficients $C_i$ are obtained from classical pre-processing, as are the one- and two-body integrals that occur in the Hamiltonian. For our work, the coefficients correspond to the two reference determinants considered: $\ket{110000110000}$ (the Hartree-Fock (HF) determinant) and $\ket{101000101000}$ (determinant generated by a double excitation from the highest occupied molecular orbital (HOMO) to the lowest unoccupied molecular orbital (LUMO)). Each entry in a ket denotes the occupancy of the spinorbital; the indexing of spinorbitals is following the big-Endian convention (order of counting qubits is $\ket{0_10_20_3\cdots}$, for instance). A schematic of the two determinants in the orbital notation is presented in Figure \ref{fig:1}(c). Furthermore, we follow the block spin notation, where the spins are arranged as $\alpha \alpha \cdots \beta \beta \cdots$. The coefficients were generated following a multiconfiguration self-consistent field (MCSCF) calculation with two MOs (HOMO and LUMO swapped) and two electrons, using the GAMESS-US quantum chemistry software \cite{GAMESS}. The one- and two-electron integrals were fetched from the DIRAC22 program \cite{DIRAC22}. As mentioned earlier, we developed our own code to remove redundancies by comparing the output states that result from the action of excitation operators on the considered reference determinants. The VQE simulations are carried out using Qiskit 0.39.5 \cite{Qiskit2021}. 

Input state preparation is an important distinguishing step between single and multireference VQE computations. For our calculation, we found an efficient isometry with only three two-qubit gates to prepare the input state (see (Figure \ref{fig:1}(d)), $C_1\ket{110000110000}+C_2\ket{101000101000}$ with $C_1 = 0.911174$ and $C_2= -0.412020$ (rounded off to six decimal places). We briefly present the intuition involved in building the circuit. We begin by preparing the state corresponding to the second term, that is, $\ket{101000101000}$, via application of three $X$ gates to the initial state $\ket{000000000000}$. We then apply a rotation gate $R_y(\theta)$ to $\ket{0_1}$ to obtain $\cos(\frac{\theta}{2})\ket{101000100000} + \sin(\frac{\theta}{2})\ket{111000100000}$. Finally, with three CX- gates (as shown in Figure \ref{fig:1}(d)), we build the state $\cos(\frac{\theta}{2})\ket{101000101000} + \sin(\frac{\theta}{2})\ket{110000110000}$. The value for $\theta$ is evaluated as $2\pi - 2\tan^{-1}\abs{\frac{C_1}{C_2}} = 3.990933$ radians. We also note that the last three qubits remain in $\ket{000}$ throughout for our choice of the two reference determinants. This state preparation circuit is a many-fold reduction over Qiskit's in-built isometry routine for the same state), however it is not easy in general to prepare an entangled state built as a linear combination of many determinants, and thus preparing a multideterminantal state for an MRUCCSD-VQE (or for that matter, quantum phase estimation \cite{kitaev1995quantum, aspuru2005simulated} or the HHL algorithms \cite{harrow2009quantum, baskaran2023adapting}) computation is an open problem in the field. 

\begin{table*}[t]
\centering
    \begin{ruledtabular}
    \caption{\label{clique_table} Contribution to the energy (in units of Hartree) from different cliques. Each clique energy is obtained by evaluating the expectation value of Hamiltonian terms in the clique with respect to $\ket{\Psi_r(\tilde{\Theta}^*)}$ which is a 1- parameter state whose parameter value is found using prior MRUCCSD-VQE simulations. We combine cliques having energy difference less than $\delta$ into supercliques. The table provides the five supercliques we find for the 12-qubit BeH$_2$ problem with $\delta=0.0001$ Ha, and for our quantum hardware executions, we pick only the top three. IYZZYIIIIIII is a shorthand for $\mathrm{I\otimes Y\otimes Z\otimes Z\otimes Y\otimes I\otimes I\otimes I\otimes I\otimes I\otimes I\otimes I}$ and the operators are in little Endian notation. Furthermore, we have omitted for the sake of simplicity the `hat' on top of each Pauli operator in a string. 
    The energies are rounded off to six decimal places. }
        \begin{tabular}{cccc}
                \textrm{Superclique no.}& \textrm{Cliques in S}& \textrm{Terms}& 
                \textrm{Clique energy} \\
                \colrule
                S$\mathrm{_0}$ & {Clique 0} & {IIIIIIIIIIII, IIIIIIIIIIIZ, IIIIIIIIIIZI, IIIIIIIIIZII,$\cdots$, ZZIIIIIIIIII} & {$-3.545374$} \\
            \colrule
                S$\mathrm{_1}$  & {Clique 13} & {IIIIIIIYZXXY, IIIIIIYYIXXI, IIIIYYIIIXXI, $\cdots$, YYIIIIIIIXXI} & {$-0.006494$} \\
                    & {Clique 14} & {IIIIIIIXZYYX, IIIIXXIIIYYI, IIIXXIIIIYYI, $\cdots$, XXIIIIIIIYY} & {$-0.006494$} \\
                    & {Clique 16} & {IIIIIIIXZXXX, IIIIIIXXIIXX, IIIIXXIIIIXX, $\cdots$, XXIIIIIIIXXI} & {$-0.006494$} \\
            \colrule
                S$\mathrm{_2}$ & {Clique 5} & {IIIIIIIYZZYI, IYZZYIIIIIII, IIIIIIIYZZYZ,$\cdots$, ZYZZYIIIIIII} & {$-0.005809$} \\
                   & {Clique 6} & {IIIIIIIXZZXI, IXZZXIIIIIII, IIIIIIIXZZXZ, $\cdots$, ZXZZXIIIIIII} & {$-0.005809$} \\
                   & {Clique 79} & {IXZZXIIYZZYI}  & {$-0.005809$} \\
                   & {Clique 80} & {IYZZYIIXZZXI}  & {$-0.005808$} \\ 
            \colrule
                S$\mathrm{_3}$ & {Clique 11} & {IIIIIIIYZYYY, IIIIIIYYIIYY, IIIIYYIIIIYY, $\cdots$, YYIYYIIIIIII} & {$-0.005533$} \\   
            \colrule
                S$\mathrm{_4}$ & {Clique 12} & {IIIIIIIXZXYY, IIIIIIXXIIYY, IIIIXXIIIIYY, $\cdots$, XXIXXIIIIIII} & {$0.000961$} \\ 
                 & {Clique 101} & {IIIXXIIYZYII, IXZXIIIYZYII} & {$0.000961$} \\
                & {Clique 102} & {IIIYYIIXZXII, IYZYIIIXZXII} & {$0.000961$} 
                   
        \end{tabular}
\end{ruledtabular}
\end{table*}

\begin{table*}[t]
\centering
    \begin{ruledtabular}
     \caption{\label{mp2_table} Energy contributions (in units of Hartree) with MP2 and MP3 amplitudes from different supercliques with a threshold of $0.001$ Ha. Here each clique energy is evaluated with respect to the MRUCCSD ansatz $\ket{\Psi({\vec{\Theta}}_{M})}$   where, $\vec{\Theta}_{M}$ is the vector containing MP2 (for doubles) and MP3 (for singles) excitation amplitudes.} 
     \begin{tabular}{cccc}
        \textrm{Superclique no.}& 
        \textrm{Cliques in S$\mathrm{'}$}& \textrm{Terms}& \textrm{Clique energy} \\
        
        \colrule
        S$\mathrm{'_0}$ & {Clique 0} & {IIIIIIIIIIII, IIIIIIIIIIIZ, IIIIIIIIIIZI, IIIIIIIIIZII,$\cdots$, ZZIIIIIIIIII} & {$-3.496076$}\\
        
        \colrule
        S$\mathrm{'_1}$ & {Clique 5} &  {IIIIIIIYZZYI, IYZZYIIIIIII, IIIIIIIYZZYZ,$\cdots$, ZYZZYIIIIIII} & {$-0.009465$} \\
                        & {Clique 6} &  {IIIIIIIXZZXI, IXZZXIIIIIII, IIIIIIIXZZXZ, $\cdots$, ZXZZXIIIIIII} &{$-0.009465$} \\
                        
        \colrule
        S$\mathrm{'_2}$ & {Clique 16} & {IIIIIIIXZXXX, IIIIIIXXIIXX, IIIIXXIIIIXX, $\cdots$, XXIIIIIIIXXI} & {$-0.007449$} \\
                        & {Clique 79} & {IXZZXIIYZZYI}  & {$-0.008279$} \\
                        & {Clique 80} & {IYZZYIIXZZXI}  & {$-0.008279$} \\
        \colrule
         S$\mathrm{'_3}$ & {Clique 11} & {IIIIIIIYZYYY, IIIIIIYYIIYY, IIIIYYIIIIYY, $\cdots$, YYIYYIIIIIII} & {$-0.006981$} \\  
                        & {Clique 13} & {IIIIIIIYZXXY, IIIIIIYYIXXI, IIIIYYIIIXXI, $\cdots$, YYIIIIIIIXXI} & {$-0.007175$} \\
                        &  {Clique 14} & {IIIIIIIXZYYX, IIIIXXIIIYYI, IIIXXIIIIYYI, $\cdots$, XXIIIIIIIYY} & {$-0.007166$} \\
     \end{tabular}
\end{ruledtabular}
\end{table*} 

We now comment on our redundancy removal procedure. We begin by generating all possible active $\rightarrow$ active and active $\rightarrow$ virtual excitations from reference determinants $\ket{\Phi_1}$ and $\ket{\Phi_2}$. Furthermore, we apply $C_{2v}$ point group symmetry to reduce the number of excited determinants from each reference determinant. These excited determinants $\ket{\Psi^{(1)'}(\vec{\theta}_1')}$ and $\ket{\Psi^{(2')} (\vec{\theta}_2')}$ are compared with each other and the excitations leading to identical determinants are noted. We remove excitations leading to identical determinants from $\hat{\tau}{(\vec{\theta}_2')}$ and club the remaining excitations with $\hat{\tau}{(\vec{\theta}_1')}$. In addition to that, we merge identical excitation operators if a common excitation operator $\hat{\tilde{\tau}}$ acts on $\ket{\Phi_1}$ and $\ket{\Phi_2}$, to avoid double counting. For example, the excitation operator $0 \rightarrow 5$ occurs commonly for both the reference determinants. We only count it once.  This step is illustrated briefly in `Redundancy removal and merging excitation operators' box of Figure \ref{fig:2}. To the best of our knowledge, this is the first redundancy-free MRUCCSD-VQE approach to have been implemented for quantum hardware calculation. In a future work, we plan to replace the manual process with a theoretical framework that accounts for redundancies, such as the internally contracted MRUCC~\cite{hanauer2011pilot},  where we identify the set of all redundant excitation operators, diagonalise the overlap matrix associated to it, then use it to construct a transformation matrix which helps to go from redundant excitation operators to linearly independent excitation operators. 

\subsection{Resource reduction}\label{resred}

The MRUCCSD-VQE quantum circuit for our problem has $12243$ two-qubit gates (involves redundancy removal and merging of excitation operator). On the other hand, the Hamiltonian has $551$ Pauli terms upon Jordan-Wigner transformation. In order to execute our problem on quantum hardware and get reliable results, we need to carry out extensive resource reduction both on the wave function as well as the Hamiltonian fronts, as illustrated in Figure \ref{fig:2}. The resource reduction procedure involves several routines, all of which are carried out on a classical computer. The need for reducing the number of two-qubit gates in a circuit on a NISQ era computer can be seen using a back-of-the-envelope calculation, where with a two-qubit gate fidelity of $99.28\%$ (the fidelity observed during the time of execution of our tasks on the commercially available IonQ Forte-I quantum computer) yields an expected result fidelity of $\sim (0.9928)^{12243}=0$, whereas to obtain a result fidelity of about $0.85$, we require to optimize the circuit such that it has only about $20$ two-qubit gates. With regard to the number of Pauli words in the Hamiltonian, it is worth noting that each term would correspond to one circuit evaluation per VQE iteration, and thus to avoid accumulating errors over evaluation of several circuits as well as reduce the cost involved in such a computation, we need to employ resource reduction strategies to reduce the number of measurements. The resource reduction techniques used in the current work are based on those used in Ref. \cite{Palak2025relvqe}. 

\subsubsection{Reducing the number of two-qubit gates} 

Our resource reduction workflow begins with leveraging the $C_{2v}$ point group symmetry \cite{Cao2022pointgroup} to reduce the number of excitation amplitudes. We obtain the $C_{2v}$ symmetry-adapted excitations from each reference determinant and manually remove those excitations from $\hat{\tau}(\vec{\theta}_2^{'})$ that lead to redundant determinants. We denote the set of linearly independent amplitudes thus obtained as $\vec{\tilde{\theta}}$. We note that the two-qubit gate count reduces from $12243$ to $3747$ with the application of $C_{2v}$ symmetry while incurring no loss in the calculated ground state energy. We then perform VQE on a classical computer with our MRUCCSD ansatz to obtain the converged excitation amplitudes, $\vec{\tilde{\theta}}^*$ , which we then use to construct the circuit for $\hat{U}(\vec{\tilde{\theta}}^*)\ket{\Phi}$ and measure the Hamiltonian on the state on a quantum computer, to obtain the ground state energy. This is in contrast to a full VQE calculation, where each iteration would be carried out on a quantum computer. We do not opt for this procedure due to the prohibitively high costs involved for such a calculation, as well as the errors that would be accumulated in such a task execution. In particular, we need to execute $n_{H}n_{iter}$ number of circuits, where $n_H$ is the number of terms in the Hamiltonian, which typically scales as $N^4$ for an $N$-spinorbital calculation, and $n_{iter}$ is the number of iterations. For our work, we use the Sequential Least SQuares Programming (SLSQP) optimizer \cite{Kraft1988Seq_QP}, with which we incur $619$ iterations in a $43$-parameter VQE (as opposed to $1991$ iterations in a $141$-parameter VQE without leveraging point group symmetry). 

This step is followed by performing energy-sort VQE \cite{Fan2023es} to pick a dominant excitation. Although the original work does not advocate picking only the dominant excitation (a double excitation where the spinorbital 1 $\rightarrow$ 4 and 7 $\rightarrow$ 10; the ordering of spinorbital indices starts from $0$, and follows the block spin arrangement), this becomes a necessity in view of current-day gate fidelities. The inclusion of each additional excitation incurs many two-qubit gates, thus making the circuit optimization that follows harder and with a final two-qubit gate count that is outside of the scope of obtaining reasonable results using current hardware. The energy sort VQE step leads to $115$ two qubit gates with a loss of $1.83\%$ in the active space energy. We note that the calculated energy with this one-parameter approximation is -3.590743 Ha. Next, the one-parameter quantum circuit is optimized using Qiskit L3 \cite{Qiskit2021} $\rightarrow$ Pytket \cite{Sivarajah2021tket} $\rightarrow$ PyZX \cite{kissinger2020Pyzx} $\rightarrow$ Qiskit L3 routines in tandem. This step reduces the total two-qubit gate count to $43$ with no loss in energy. The circuit undergoes another layer of optimization by an agent trained by reinforcement learning and Graph Neural Networks to enforce rules based on ZX-calculus \cite{riu2024reinforcementlearningbasedquantum}, followed by another round of Qiskit-L3 optimization. The final optimized circuit has $20$ two-qubit gates, and we verified that the aforementioned circuit optimization steps do not lead to any loss in energy. 

\subsubsection{Reducing the number of Hamiltonian terms} 

The Hamiltonian of the system contains terms to be measured. In general, the number of terms in a molecular Hamiltonian scales as $N^4$, where we recall that $N$ is the number of spinorbitals in the problem, and thus at most as many circuits to measure. Carrying out this exercise on quantum hardware would lead to error accumulation. Hence, we group qubit-wise commuting Pauli terms in the Hamiltonian, and each such set is termed as a clique. Every term is analyzed qubit-wise and suitable gates are applied to each qubit so as to rotate the state to the shared eigenbasis of the qubit-wise commuting terms. Thus, a single unitary, $\hat{V_i}$, is constructed for a set of commuting terms. This drastically reduces the number of circuit evaluations. In this work, cliques that contribute equally to the energy are grouped under supercliques, and we measure a clique from the superclique set. This allows for fewer circuit evaluations. We find that $551$ terms get grouped into $111$ cliques, and $111$ cliques are grouped under fewer supercliques, and in particular, $5$ of them. We pick the top $3$ supercliques for our quantum hardware computations. The energy that we obtain with all of the resource reduction steps is $-3.588091$ Ha, which is less than the HF energy ($-3.570995$ Ha) by $17.096$ milliHa. It is this small value of correlation energy that we seek to capture on a NISQ era quantum computer. The details of the supercliques are given in Table \ref{clique_table}. 

\subsubsection{Alternative approaches to defining supercliques}\label{mp2sc} 

Since the selection for supercliques described in the previous sub-section is based on evaluating ${\bra{\Psi_r(\tilde{\Theta}^*)}\hat{V_i}\ket{\Psi_r(\tilde{\Theta}^*)}}$, where $\ket{\Psi_r(\tilde{\Theta}^*)}$ is a 1-parameter state whose parameter value is found using prior MRUCCSD-VQE simulations, the protocol relies on prior classical knowledge of the system, potentially undermining the benefit of using a quantum computer. To that end, we propose that in future works, alternatives such as the two possibilities we propose below be used. 

The first option relies on using a cheaper method to find a more approximate wave function rather than use the MRUCCSD-VQE one. To that end, a multireference M{\o}ller-Plesset perturbation theory to second order in energy (MP2) can be employed. However, since we are mainly interested in clubbing cliques under a superclique, which does not necessarily demand accurate energy calculations but rather only a reliable approximations to them, a viable alternative could be to: 

\begin{itemize}
\item Calculate the double excitation amplitudes ($t_{ijab}$) using MP2 and singles ($t_{ia}$) using MP3 for each reference determinant using the following expressions \cite{sugisaki2022variational}, 
\begin{eqnarray}
t_{ijab} &=& \frac{h_{ijba} - h_{ijab}}{\varepsilon_j +\varepsilon_i - \varepsilon_a - \varepsilon_b},\nonumber \\
\mathrm{and} \nonumber\\ 
t_{ia} &=& \frac{2(\sum_{jbc} t_{ijbc} h_{jabc} - \sum_{jkb} t_{jkba} h_{jikb})}{\varepsilon_i - \varepsilon_a}\nonumber,
\end{eqnarray}
where $\varepsilon_i, \varepsilon_j, \varepsilon_a, \varepsilon_b$ are respectively the orbital energies of the occupied orbitals $i$ and $j$ and the virtual orbitals $a$ and $b$. A partial renormalization with size-consistency of singles amplitudes through $ t_{ia}(scaled) = \frac{t_{ia}}{1+\sum_{d} (t_{id})^2}$, has proven to accelerate the convergence of the VQE \cite{MOCHIZUKI2005165}. For that reason, we have implemented this scaling in our calculations. In the above expression, the index $d$ runs over all virtual orbitals. We also note that MP2 scales as $N^4$, and thus is only as expensive as a HF procedure. 
\item After obtaining the estimates for t-amplitudes associated with both the reference determinants (which we also scale by their respective coefficients), which we denote as $\{t_1\}$ and $\{t_2\}$ for MP2 and MP3 calculations involving the unperturbed states $\ket{\Phi_1}$ and $\ket{\Phi_2}$ respectively. We then remove redundancies in the same way we described it earlier in Section \ref{mrucc} and merge the two lists while ensuring that in case a linearly independent amplitude corresponding to the first determinant and another corresponding to the second share the same indices, we replace the two amplitudes by a new one that is their weighted sum, weighted by their respective coefficients $C_1$ and $C_2$, 
\item Construct the MRUCCSD-VQE quantum circuit with these amplitudes as parameters and get expectation values for each of the cliques, 
\item If the energies corresponding to some cliques are different by some set threshold, $\delta$, then club them together under a superclique. 
\end{itemize} 

We now report our results using this approach for the 12 qubit BeH$_2$ problem that we consider. Since we are only proposing the method and are not testing this on quantum hardware, we consider all {43} parameters for the calculation. Table \ref{mp2_table} presents our results, and lists for convenience only those supercliques with energy greater than $0.005$ Ha. The energies correspond to the UCCSD energies with the MP2 and MP3 amplitudes. The error in UCCSD-VQE active space energy with 1 parameter by measuring top 3 supercliques (S$'_0$, S$'_1$ and S$'_2$), obtained from MP2 calculation, is $0.44\%$ while the error incurred if we had obtained top 3 supercliques by classical VQE simulation (S$_0$, S$_1$ and S$_2$) is $0.074\%$.  With the improved gate fidelities, one can increase the number of supercliques to be measured on hardware resulting in the  reduction of error in the total energy. 

Another alternative would be to opt for a wave function-independent way. For example, recalling that the qubit operator Hamiltonian is of the form $H = \sum_{\alpha} h_\alpha P_\alpha$, where $P_\alpha$ is a Pauli word, we could pick the largest coefficient, $h_\alpha$, from the elements of each clique, and then sort these coefficient values in descending order to set a threshold and pick the top $\mathcal{L}$ dominant coefficients. We note that this does not involve supercliques, but rather a simple form of screening to select Hamiltonian terms for a quantum hardware calculation. 

\section{Results and Discussions}\label{results} 

\subsection{Numerical results} 

\begin{figure}[t]
\centering
\setlength{\tabcolsep}{1mm}
\begin{tabular}{c}
\hspace{-0.5cm}
\includegraphics[height=90mm,width=100mm]{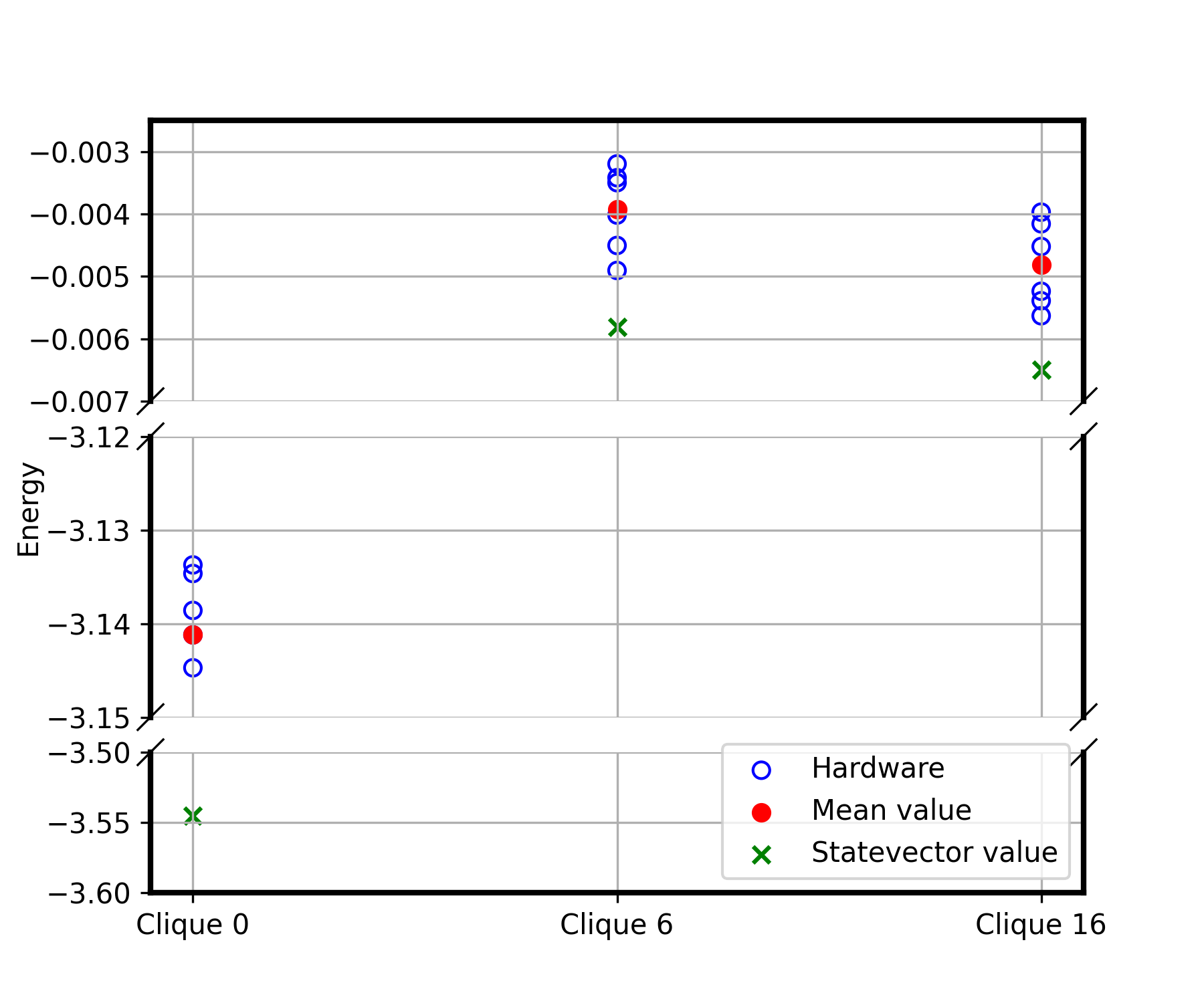} 
\vspace{-0.5cm}
\end{tabular}
\caption{Quantum hardware results (with error mitigation) for the ground state energy (in units of Ha) using our resource optimized MRUCCSD-VQE circuit with optimal parameters and by considering three supercliques. } 
\label{fig:3}
\end{figure}

We first begin with the computational settings of the IonQ Forte-I hardware during the time of executing our target jobs: Average one-qubit gate fidelity: $99.98\%,$ average two-qubit gate fidelity: $99.28\%$, and readout fidelity: $99.17\% $. We note here that the reason for concerning ourselves more with two-qubit gate count over the one-qubit ones is due to the fact that the two-qubit gate fidelities are the lowest when compared with one-qubit gate or readout fidelities, and thus have a big impact on the final results. Furthermore, $T_1$ = 100 seconds, $T_2$ = 1 second (not to be confused with the coupled cluster excitation operators $\hat{T_1}$ and $\hat{T_2}$), one-qubit gate duration = 130 microseconds, two-qubit gate duration = 970 microseconds, and readout duration = 150 microseconds. For all of our calculations, we use $4000$ shots, include error mitigation (debiasing \cite{maksymov2023symmetry}), and for the reported results in this work, we average over $5$ repetitions for the dominant clique (clique $0$), and $6$ for the other two (cliques $6$ and $16$). We also note that, the error in measurement of clique $0$ is substantially larger than the error in measurement of clique $6$ and clique $16$ indicating the presence of relative error from the hardware side which seems to be dependent on the magnitude of value to be captured. See Figure \ref{fig:3} for reference. 

 Upon executing the aforementioned tasks on the Forte-I device, we implement a post-selection strategy for clique $0$, where only the bitstrings that conserve the particle number are retained in our final results. We note that this technique can be applied to Pauli terms which are only in computational basis \cite{Going2023mol_vqe}. Figure \ref{fig:3} represents the active space energy after error mitigation (for all cliques) and post-selection (only for clique $0$). The total energy in the active space is calculated as $E =\sum_{\alpha \in Cl_0}h_{\alpha}\expval{\alpha} + 4 \sum_{\beta \in Cl_6}h_{\beta}\expval{\beta} + 3\sum_{\gamma \in Cl_{16}}h_{\gamma} \expval{\gamma}$. The error in the active space energy from the hardware is $11.61 \%$ with respect to the expectation value of same Hamiltonian evaluated using the statevector backend. When we account for the contributions from nuclear repulsion and core energies, the error in total energy is $2.69\%$. To obtain the total energy, we need to add the contributions from nuclear repulsion and core orbitals. The energy due to nuclear repulsion is given by $\sum_{A=1}^{M} \sum_{B>A}^M \frac{Z_AZ_B}{R_{AB}}$, where $M$ is the total number of nuclei in the system and $R_{AB}$ is the distance between the nuclei $A$ and $B$. The energy contribution from the core orbitals is $\braket{i|h|i}+\sum_{i \ne j} \braket{ij||ij}$, where $\braket{i|h|i} = \int \phi_i^{*}(\vec{r}_1) \bigg(-\frac{1}{2} \nabla_1^2-\sum_A \frac{Z_A}{|\vec{r}_1 - R_A|}\bigg) \phi_i(\vec{r}_1) d\vec{r}_1$ and $\braket{ij||ij} = \int \phi_i^{*}(\vec{r}_1) \phi_j^{*}(\vec{r}_2) \frac{1}{|\vec{r}_1-\vec{r}_2|} \phi_i(\vec{r}_1) \phi_j(\vec{r}_2) d\vec{r}_1 d\vec{r}_2 - \int \phi_i^{*}(\vec{r}_1) \phi_j^{*}(\vec{r}_2) \frac{1}{|\vec{r}_1-\vec{r}_2|} \phi_i(\vec{r}_2) \phi_j(\vec{r}_1) d\vec{r}_1 d\vec{r}_2$. We note that $\vec{r}_1$ and $\vec{r}_2$ denote the positions of the electron and $R_A$ denotes the position of the nucleus \cite{szabo1996modern}.The total energy itself is $-15.052701$ Ha (the HF energy is $-15.452333$ Ha, the one-parameter VQE with dominant parameter yields $-15.472081$ Ha, energy after accounting for only three dominant supercliques is $-15.469429$ Ha, and the full VQE energy is $-15.538922$ Ha). However, when we go beyond the precision in the total ground state energy and check the amount of correlation energy that our computation has captured, that is, $E-E_{HF}$, where $E$ is the computed energy with the reduced wave function and Hamiltonian and $E_{HF}$ is the HF energy calculated classically, we find it to be very limited at best, due to noise. In fact, recalling that the correlation energy that we sought to capture is $-17.096$ milliHa, we find that our calculation captures $400$ milliHa but on the other side of the HF value. An earlier work that calculates ground state energies using the UCCSD-VQE approach on quantum hardware also report similar findings (large errors in correlation energy itself, although the total energy error is small due to the large HF contribution to it) for their 6- and 12-qubit computations on the Aria-I and the Forte-I quantum computers respectively \cite{Palak2025relvqe}, but we find that in our MRUCCSD-VQE case, the issue is more pronounced. In fact, the clique $0$ contribution to the energy is $-3.545374$ Ha, whereas the HF energy is $-3.570995$ Ha in our case. However, in the UCCSD-VQE case that the authors in Ref. \cite{Palak2025relvqe} considered, the dominant clique subsumed the HF energy in it, and thus measuring one clique was sufficient in their case to obtain a total energy lower than the HF value. We note that a larger active space choice and/or a lower degree of approximation (for example, picking more parameters from energy sort VQE) would have led to a larger amount of correlation energy to capture, but would have run into the issue of very deep circuit to evaluate in a noisy setting. Thus, we conclude that we need better resilience to noise on the quantum hardware front to be able to capture the right correlation energy trend in the VQE framework. We will now quantify this statement with back-of-the-envelope estimates. In order to reliably capture the correlation energy, we estimate the required two-qubit gate fidelity to be $0.99999$ or 5 9s. We arrive at the estimate as follows: We recall that the total energy to be captured in our active space is $-3.590742$ Ha and the total correlation energy is $19.747$ milliHa ($-17.096$ milliHa is the correlation energy that we sought to capture, given that we selected only the top three supercliques; here we lift that restriction). Thus, the percentage fraction of correlation energy is about $0.55$, that is, even to capture HF energy reliably, we need an expected result fidelity of $0.9945$. We assume that we aim at a target result fidelity of $0.9999$, so that the correlation energy thus captured is $-19.388$ milliHa (this is a sizeable chunk of the total correlation energy which we ideally would like to capture). In such a case, we already see that with 20 two-qubit gates (which we recall was a result of employing significant amount of optimization strategies), the required two-qubit gate fidelity is $0.99999$ or 5 9s. For perspective, the best known quantum computers as of the time of completion of this work have achieved two-qubit gate fidelities of 2 9s. If we further extend the same analysis to get an idea of the required two-qubit gate fidelities to execute a VQE with all parameters for our target molecule and expect to reach a result fidelity of 4 9s, we require a two-qubit gate fidelity of 7 9s. Since it is impractical to expect this much of two-qubit gate fidelity in the coming few years, one needs to incorporate the circuits with quantum error correction to reduce the logical error rate \cite{m3}. 

\subsection{Scaling} 

We now comment on scaling. Several works have commented on the scaling behaviour in variational algorithms (for example, see Refs. \cite{Jules2022VQE_review, zade2024capturing,nphard}; the prospects of an advantage from VQE seem unclear). In this work, we qualitatively comment on the additional classical and quantum resources required for our MRUCCSD-VQE computation over the traditional UCCSD-VQE implementation: input state preparation (quantum computing step) and redundancy removal (classical computing step). We also comment very briefly on the expected scaling of resource reduction and superclique selection steps of our workflow. Throughout, we focus on worst-case scaling (upper bounds). 

The input state preparation in the traditional UCCSD-VQE method is trivial, as it only involves one-qubit gates. In contrast, for an MRUCCSD-VQE approach, this step is hard (for example, see Ref. \cite{zade2024capturing}, where the authors consider a simple example using Qiskit's isometry routine to analyze the cost from this step, and find it to be exponential). It is worth adding that there are works in literature that attempt to address this open problem of input state preparation cost for multi-configurational states in the context of other algorithms (for example, see Refs. \cite{multiref_1, multiref_2, Veis2014, Norm2018, Erakovic2025}). 

In our implementation, the redundancy removal step would add an extra classical cost of $\sim N^4$ for an MRUCCSD calculation, where $N$ is the total number of spinorbitals. We recall that we prepare a list of excited determinants when the relevant excitation operators act on the first reference determinant, with the length of the list being $\sim N^2$. We do the same for the second reference determinant to generate a list of length $N^2$. It is not $\sim n_o^2n_v^2 \sim N^4$ but rather only $N^2$, because we only need to consider those occupied spinorbitals that are in the active space that generates the reference determinants, and we assume that the number of occupied spinorbitals in this space grows extremely slowly (to the extent of approximating it to be near-constant) in view of a user employing their `chemical intuition' to decide on the space. Any excitation resulting from occupied spinorbitals outside this set will never give rise to a redundancy, since the resulting excited determinant will never match with any  excited determinant generated from excitations arising from the other reference determinant. We now compare the two lists, each of length $N^2$, and thus the entire process takes $\sim N^4$ time. 

We briefly comment on our resource reduction steps, as this step is essential to make possible a quantum hardware execution even for a small molecule in a small active space. We expect that the RL-based ZX calculus step is costlier than the rest in view of training costs involved. Ref. \cite{riu2024reinforcementlearningbasedquantum}, which introduced the method, assesses the cost to be at most sub-exponential. This can still be costly in practice. The superclique selection adopted for our quantum hardware runs presupposes a classical VQE simulation. In view of this limitation, we instead address the scaling of the MP2 approach that we introduce in Section \ref{mp2sc}. All steps involved here are classical, except the evaluation of MRUCCSD-VQE energy with MP2 and MP3 amplitudes on a quantum circuit. The number of two-qubit gates in the UCCSD circuit with MP2 and MP3 amplitudes depends on the number of excitation operators which scales as $\sim N^4$. Then, the number of Pauli gadgets in the circuit also scales as $\sim N^4$. Hence, the two qubit gate count in the UCCSD circuit would scale as $\sim N^5$ assuming each Pauli gadget would require at most $N$ two qubit gates. Here, we assume the Jordan-Wigner transformation. If Bravyi-Kitaev transformation is used, the number of two-qubit gates for each excitation/de-excitation operator scales as $\mathcal{O}(log(N))$. We note that the cost of initial state preparation is excluded from the analysis. Now, the cost of finding cliques depends on the number of Pauli terms in the Hamiltonian which is at most $N^4$. The simplistic brute force approach of checking for $[H_i,H_j]_q$ (where the subscript denotes qubit-wise commutation) for every possible value of $j$, and then repeating this exercise for each $i$ (which gets successively cheaper due to having to evaluate fewer commutations) means that the cost of finding cliques comes to $\sim N^5$. We note that the matrix multiplication steps themselves are constant in their cost, since they always involve two $2 \times 2$ Pauli matrices being multiplied. We now move to the superclique selection step. Finding the t-amplitudes via MP2 is at most $\sim N^4$, and so is the cost of redundancy removal as seen in the earlier paragraphs. Thus, the superclique selection procedure is dominated by the clique finding step, which in turn scales as $N^5$. 

\section{Conclusion} 

In conclusion, we carry out a 12-qubit multireference unitary coupled cluster VQE calculation on a trapped ion quantum hardware to obtain the energy of the BeH$_2$ molecule in a geometry where the role of strong correlation effects is significant. We have ensured a redundancy-free calculation by removing repeating determinants. The limitations imposed by current-day quantum hardware demanded the use of resource reduction techniques to reduce the two-qubit gate count and the number of measurements, besides necessitating the use of error mitigation (debiasing) and post-selection (based on particle number conservation). By leveraging symmetry, using energy sort VQE, and using pipeline-based quantum circuit optimization, we reduced the two-qubit gate count from $12243$ to $20$ with a loss of $0.45\%$ in the total energy. We used the notion of supercliques, where we partition a Hamiltonian into sets of qubit-wise mutually commuting terms (cliques), and then combine cliques yielding the same energy contributions into supercliques. Picking only the important supercliques leads to a drastic reduction in the number of circuits to evaluate on quantum hardware. Furthermore, we use a simple isometry, which leads to an input state preparation circuit with only $3$ two-qubit gates. We also note that in view of the prohibitive costs and accumulation of errors, we prepare a circuit with optimized parameters obtained from VQE simulation and then measure the Hamiltonian on that prepared state on a quantum computer. We find that the error in ground state energy obtained on the Forte-I quantum computer relative to that evaluated on a traditional computer with the same reduced problem setting is only $2.69\%$. However, although employing a series of resource reduction techniques significantly lowers the depth of the quantum circuit while not losing a notable fraction of the energy to be captured, a combination of the small magnitude of correlation energy in the chosen active space and the high physical error rates in current-day quantum computers yields an energy that is still larger than the HF value. We expect that with further advances in the quantum hardware front, one can carry out multireference UCCSD-VQE computations with better precision. 

\section{Acknowledgments} 

The work was carried out as a part of the Meity Quantum Applications Lab (QCAL) Cohort 2 projects. VSP acknowledges support from CRG grant (CRG/2023/002558). VSP, PC, and DS acknowledge Prof. Bhanu Pratap Das for initial discussions on multireference coupled cluster theories, Ms. Aashna Anil Zade for fruitful discussions on various concepts, Dr. Cedric, Dr. Mao, and Mr. Jeffrey without whose support in fixing a bug at the last minute would have made quantum hardware execution a huge challenge, Dr. Subimal Deb for deliberations on single particle basis sets, and Mr. Sudhindu Bikash Mandal for support with AWS Braket. KS acknowledges support from Quantum Leap Flagship Program (Grant No. JPMXS0120319794) from the MEXT, Japan, Center of Innovations for Sustainable Quantum AI (JPMJPF2221) from JST, Japan, and Grants-in-Aid for Scientific Research C (21K03407) and for Transformative Research Area B (23H03819) from JSPS, Japan. JR acknowledges support from the  Agència de Gestió d’Ajuts Universitaris i de Recerca through the DI grant (No. 2020-DI00063). JR and JN acknowledge support from MICIU/AEI/10.13039/501100011033/ FEDER, UE.

\section{Data availability} 

All the data is included in the article. The details of our code will be available on reasonable request. 

\bibliography{apsbib}

\end{document}